\documentclass[sigconf]{acmart} 
\AtBeginDocument{%
  }
\usepackage{hyperref}
\usepackage{tablefootnote}

\copyrightyear{2025}
\acmYear{2025}
\setcopyright{rightsretained}
\acmConference[DLfM 2025]{12th International Conference on Digital Libraries for Musicology}{September 26, 2025}{Seoul, Republic of Korea}
\acmBooktitle{12th International Conference on Digital Libraries for Musicology (DLfM 2025), September 26, 2025, Seoul, Republic of Korea}
\acmPrice{}
\acmDOI{10.1145/3748336.3748341}
\acmISBN{979-8-4007-2083-3/25/09}

\begin{document}

\title{The IRMA Dataset: A Structured Audio–MIDI Corpus for Iranian Classical Music}


\author{Sepideh Shafiei}

\orcid{https://orcid.org/0000-0002-8214-6024}
\authornotemark[1]
\affiliation{%
  \institution{Cu Test Inc.}
  \city{Berkeley}
  \state{California}
  \country{USA}
}

\author{Shapour Hakam}
\affiliation{%
  \institution{Cu Test Inc.}
  \city{Berkeley}
  \state{California}
  \country{USA}}



\begin{abstract}

We present the \textbf{IRMA Dataset} (\textbf{I}ranian \textbf{R}adif \textbf{M}IDI \textbf{A}udio), a multi-level, open-access corpus designed for the computational study of Iranian classical music, with a particular emphasis on the \textit{radif}—a structured repertoire of modal-melodic units central to pedagogy and performance. The dataset combines symbolic MIDI representations, phrase-level audio–MIDI alignment, musicological transcriptions in PDF format, and comparative tables of theoretical information curated from a range of performers and scholars. We outline the multi-phase construction process, including segment annotation, alignment methods, and a structured system of identifier codes to reference individual musical units. The current release includes the complete \textit{radif} of Karimi; MIDI files and metadata from Mirzā~Abdollāh’s \textit{radif}; selected segments from the vocal \textit{radif} of Davāmi, as transcribed by Pāyvar and Fereyduni; and a dedicated section featuring audio–MIDI examples of \textit{tahrir} ornamentation performed by prominent 20\textsuperscript{th}-century vocalists. While the symbolic and analytical components are released under an open-access license (CC BY-NC 4.0), some referenced audio recordings and third-party transcriptions are cited using discographic information to enable users to locate the original materials independently, pending copyright permission. Serving both as a scholarly archive and a resource for computational analysis, this dataset supports applications in ethnomusicology, pedagogy, symbolic audio research, cultural heritage preservation, and AI-driven tasks such as automatic transcription and music generation. We welcome collaboration and feedback to support its ongoing refinement and broader integration into musicological and machine learning workflows.

\end{abstract}




\maketitle

\begin{center}
\small \textit{Author’s version (preprint). This paper appears in the Proceedings of the Digital Libraries for Musicology (DLfM) 2025. 
Please cite the published version. DOI: \texttt{10.1145/3748336.3748341}}
\end{center}
\section{Introduction}

This paper presents the development of a computationally structured and annotated Audio-MIDI dataset for Iranian classical music, with a focus on the \textit{radif}, a central repertoire consisting of hundreds of modal and melodic pieces known as \textit{gushes}, organized into collections of \textit{dastgāhs} and \textit{āvāzes}. While transcriptions of \textit{radifs} have been published and are publicly available, no previous dataset has included MIDI files that capture performance nuances—such as pitch bends and microtonal variation—alongside aligned symbolic data in a format suitable for computational analysis.  The construction of this dataset began in 2016 and has involved multiple stages of music engraving, transcription alignment, and performance analysis. A number of trained music engravers contributed to the digitization and typesetting of available transcriptions, ensuring both musical accuracy and fidelity to original sources. This long-term collaborative effort reflects a commitment to both cultural preservation and computational accessibility. Prior to its formal release as the IRMA Dataset, parts of the corpus\footnote{This portion of the corpus was released in 2021 as a GitHub repository and includes the MIDI files of all 249 gushes of Mirzā Abdollāh's radif: \url{https://github.com/SepiSha/radif-Midi}.}
 were already used in computational studies of Iranian classical music \cite{Khalij}. 

The IRMA dataset was initially constructed by primarily focusing on the vocal \textit{radif} of Karimi (145 \textit{gushes}), combining high-quality audio with three layers of transcription, ranging from published notation to performance-aligned symbolic files. The IRMA dataset is further enriched by parallel materials from other \textit{radifs}, including those of Mirzā~Abdollāh (249 \textit{gushes}) and Davāmi (188 \textit{gushes}), and is organized across multiple levels of processing, from raw audio to computationally extracted features such as pitch histograms, pitch drift diagrams, and sentence-level alignments. IRMA dataset also includes a dedicated section on \textit{tahrir}, a characteristic vocal ornamentation in Iranian classical music \cite{S-12,Mir-1}. This section contains over two hundred annotated samples of \textit{tahrir} phrases, each accompanied by symbolic transcriptions and MIDI files, drawn from performances by various prominent singers of the 20\textsuperscript{th} century. The dataset also includes multiple tables compiling theoretical perspectives \cite{M1, T-33, Al-1,xos} by prominent Iranian musicologists on the range and distinguishing features of each \textit{gushe}, presented in a standardized and comparable format.

Designed with both musicological depth and technical interpretability in mind, the IRMA dataset supports research in symbolic audio analysis, performance modeling, and culture-specific computational musicology. This work aligns with broader efforts in digital musicology and cultural heritage preservation by contributing a structured, computationally accessible corpus of a historically transmitted repertoire. By digitizing, encoding, and aligning audio and symbolic representations of the \textit{radif}, the dataset supports MIR-driven analysis while preserving culturally specific knowledge in a form accessible to both researchers and musicians.


\section{Background and Related Work}
Previous computational research on the \textit{radif} has addressed topics such as microtonal intervals~\cite{Sanati,S-6,S-11,Poorhaydari}, vocal ornamentation~\cite{S-12,Mir-1}, and mode classification and recognition~\cite{Hey1,Luciano}. Recent efforts have also begun to develop corpora for Iranian music, including the Persian Piano Corpus~\cite{Rasouli}, which focuses on instrument-based feature extraction in relation to \textit{dastgāh} structure, and KDC~\cite{BN-1}, which provides an open corpus for the analysis of \textit{dastgāhi} music. These datasets represent valuable steps toward enabling computational research on Persian music. However, IRMA is the first dataset to include both MIDI representations of \textit{radifs} and symbolic-audio alignment, along with detailed treatment of vocal ornamentation and phrase-level performance analysis. By focusing primarily on solo voice and \textit{ney}\footnote{An end-blown flute widely used in Iranian classical music.}, our dataset enables highly accurate pitch recognition and supports clean, reliable audio-symbolic alignment.

Moreover, this is the first dataset of its kind to include the complete \textit{radifs} of prominent masters. It offers a multi-level structure that integrates aligned audio and symbolic data, pitch-bend and microtonal information, and dedicated resources for analyzing vocal techniques such as \textit{tahrir}, thereby providing a comprehensive platform for both MIR and musicological applications. In addition, the dataset incorporates extensive theoretical material curated from multiple transcriptions and musicological sources, supporting efforts to better understand—and potentially bridge—the gap between theoretical frameworks and performance practice in this tradition.

\section{Overview of the IRMA Dataset}

\begin{figure}[h]

  \centering
  \includegraphics[width=\linewidth]{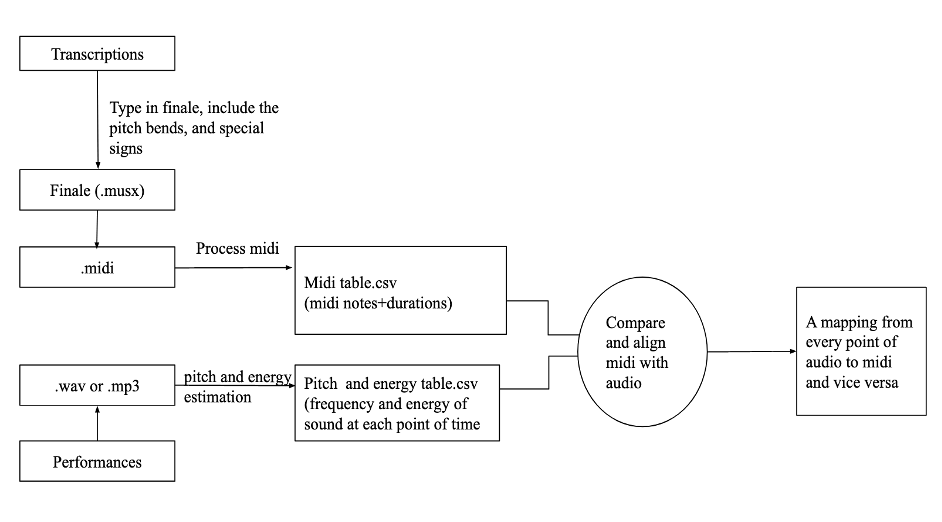}
  \caption{Pipeline for generating Audio-Midi alignment}
  \label{overview}
  \Description{An Overview of the Data}
\end{figure}

The development of the IRMA database involves multiple stages. In this section, we outline the foundational structure and key components of the process. Figure~\ref{overview} provides an overview of the initial steps in the dataset’s construction.

\begin{figure}[h]

  \centering
  \includegraphics[width=\linewidth]{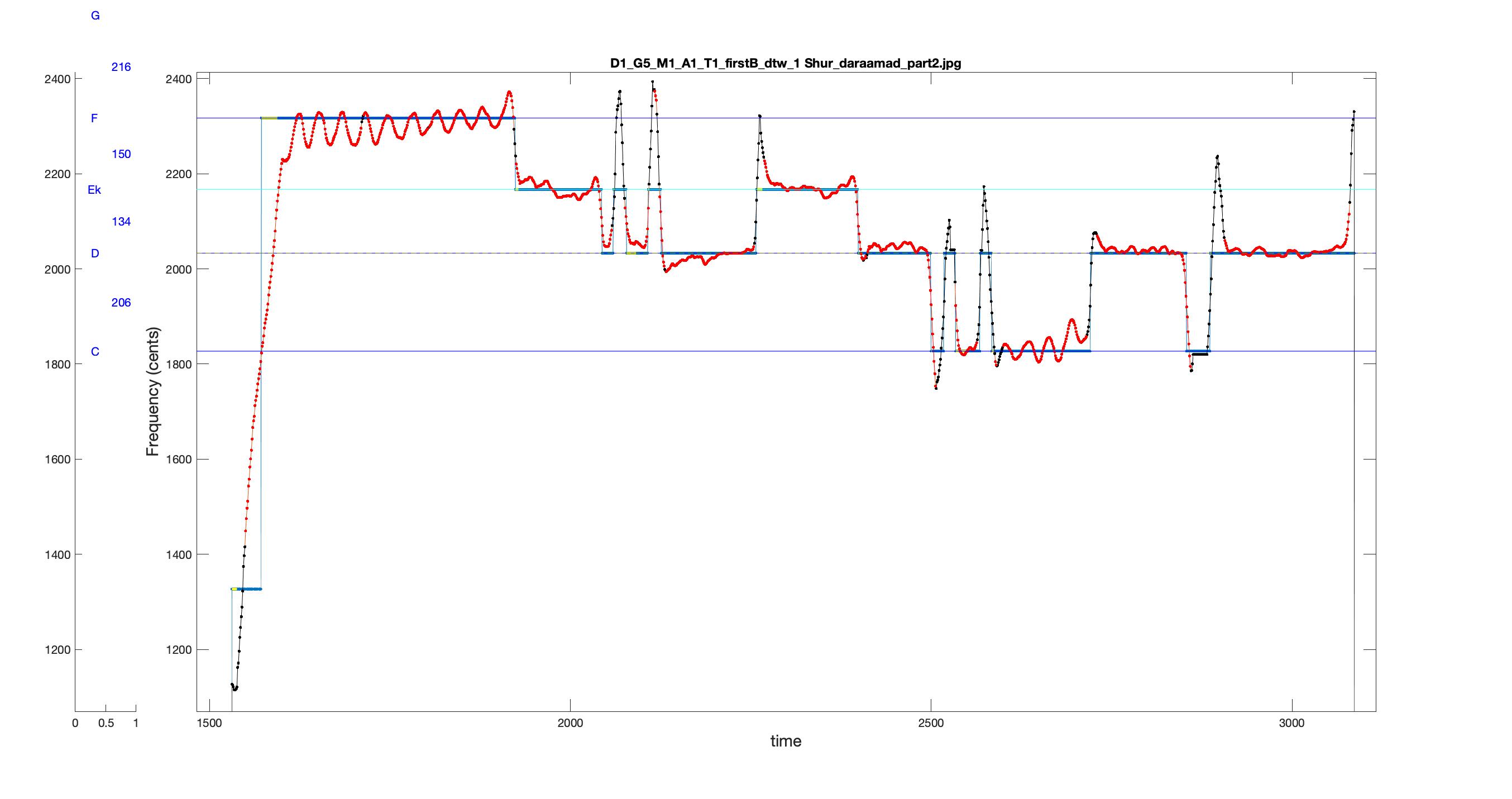}
  \caption{Audio-Midi Alignment in the dataset}
  \label{alignment}
  \Description{Audio-Midi Alignment in the dataset}
\end{figure}

The IRMA dataset includes a full data and analysis of Karimi’s \textit{radif}, the MIDI files for Mirzā Abdollāh’s \textit{radif} based on Talāyi’s transcription, as well as selected segments of Davāmi’s vocal \textit{radif}, transcribed by Pāyvar and by Nimā Fereyduni. More information and details are provided in Table~\ref{tab:radif_repertoires}.  Overall, the dataset is organized into three levels of data:

\begin{itemize}
    \item \textbf{Raw data:} This includes MP3 or WAV recordings of performances, along with the original transcriptions of available \textit{radifs}.
    
    \item \textbf{First-level processed data:} These consist of transcriptions retyped in Finale software, with added pitch bends and special notational signs from the original sources. This level also includes CSV files for pitch recognition and RMS energy analysis of the performances, as well as the corresponding MIDI and PDF files generated from Finale. A sample transcription available in the IRMA dataset is shown in Figure \ref{transcription}. In this notation, the p-shaped sign on the note E represents \emph{koron}, a microtonal modification in Iranian Classical Music that lowers the pitch by approximately a quarter-tone, though its exact tuning varies depending on performance context.
    
  \begin{figure}[h]

  \centering
  \includegraphics[width=\linewidth]{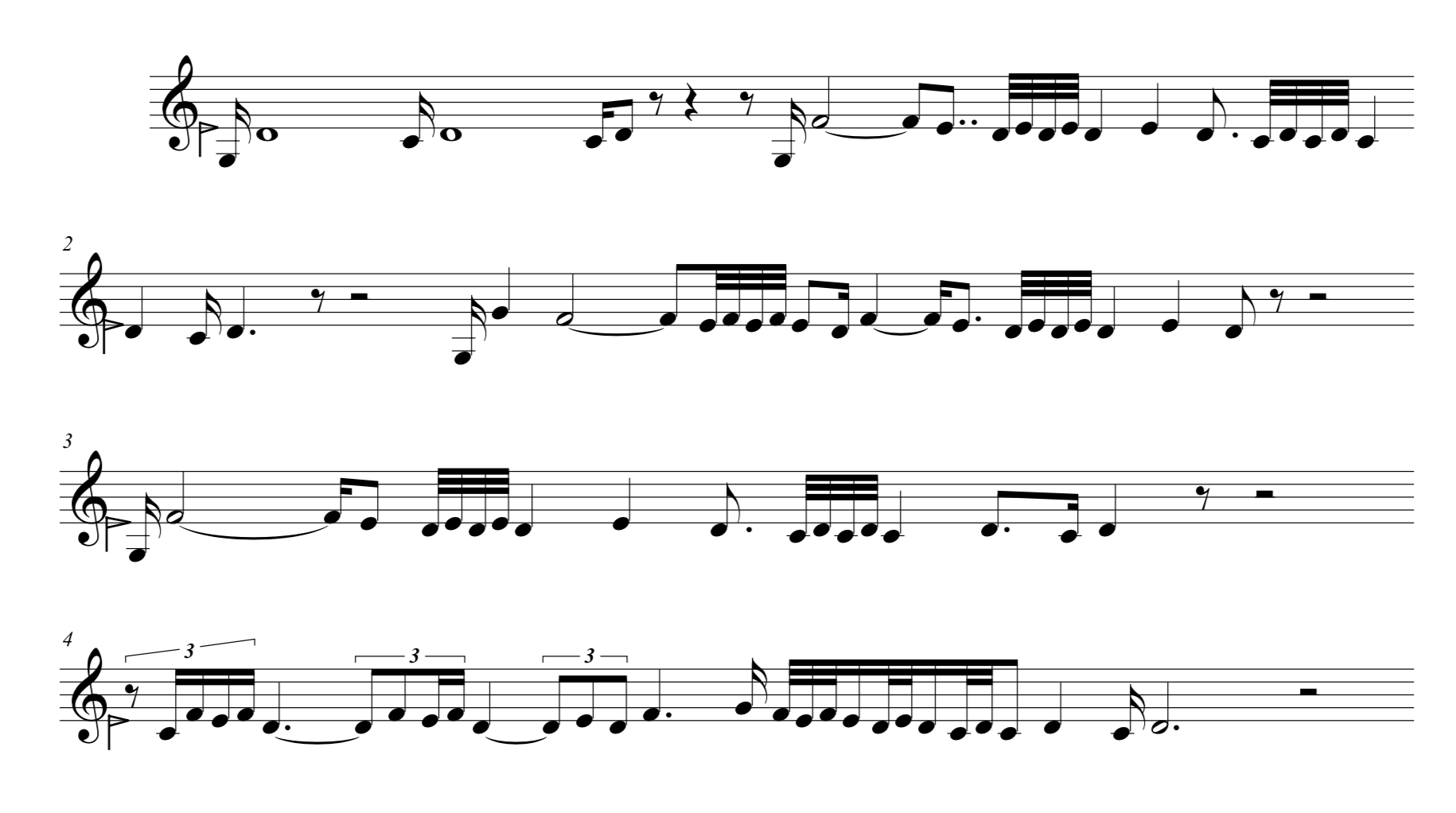}
  \caption{Transcription of Karimi's Darāmad of Shur, typeset in Finale}
  \label{transcription}
  \Description{Transcription of Darāmad of Shur in Finale}
\end{figure}
    
    \item \textbf{Second-level processed data:} This level comprises the outputs of our computational analysis and musicological research. The data—saved in TXT, CSV, and image formats—include audio and MIDI pitch histograms, pitch drift diagrams, sentence-by-sentence alignments between audio and MIDI, note-by-note comparisons of audio-derived timing and notated MIDI duration, and phrase segmentation ranges. Examples of aligned sentence diagrams are shown in Figure~\ref{alignment}.  
\end{itemize}

\begin{figure}[h]

  \centering
  \includegraphics[width=\linewidth]{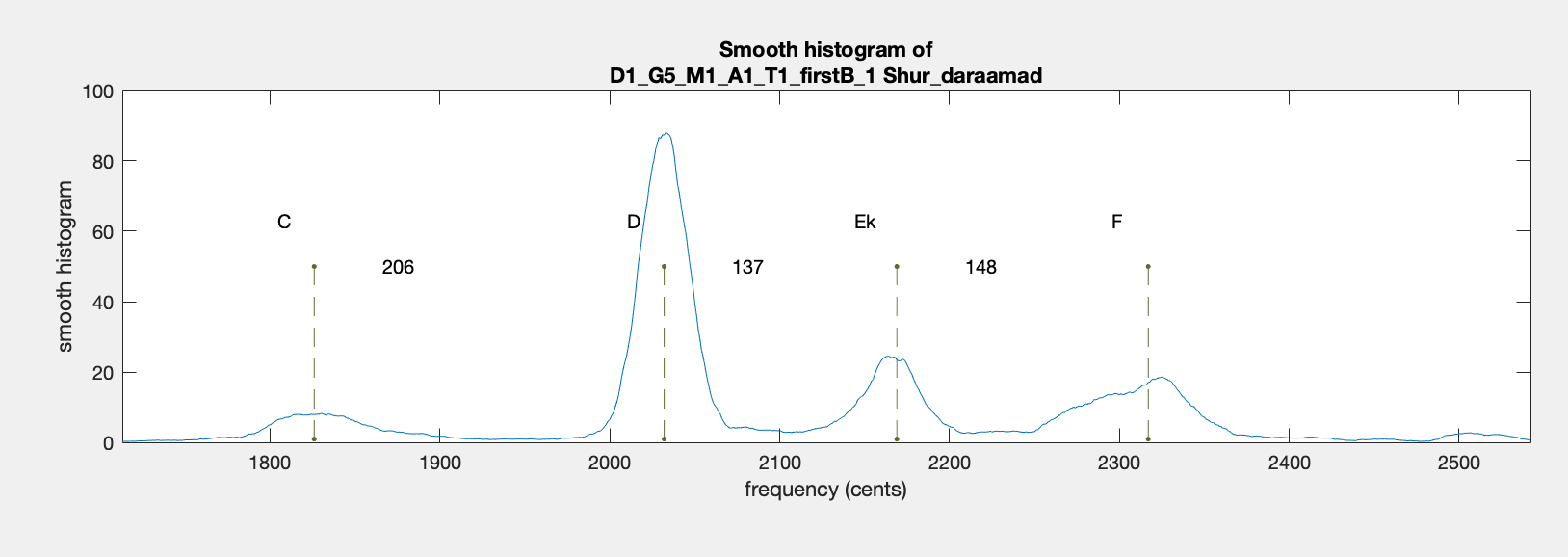}
  \caption{Audio Histogram of a Piece in Dataset}
  \label{histogram}
  \Description{Audio Histogram of a Piece in Dataset}
\end{figure}

Figure~\ref{histogram} shows a sample audio-based pitch histogram for a selected \textit{gushe} from Karimi’s \textit{radif}. In addition to the histogram itself, the graph includes the precise intervals between consecutive notes within the range of the piece. Providing this information in both tabular and graphical form for each piece enhances the analytical value of the dataset for researchers and musicians alike. Pitch histograms have been widely used in the MIR literature for analyzing musical intervals~\cite{B-10, S-13, K-14}.

In a separate table within the dataset, we have also included \textbf{theoretical models} and analytical descriptions by various Iranian musicologists and theorists\footnote{A portion of this was released in 2023 as a GitHub repository and includes the 26 tables related to the theoretical works of Talāyi and Jafarzādeh.  \url{https://github.com/SepiSha/radif-theoretical-tables}.}. These materials provide information such as the expected range, tetrachord and pentachord components of each gushe, and functional notes (e.g., \textit{shāhed}, \textit{ist}, \textit{moteghayyer}\footnote{\textit{Shāhed} (“witness”) refers to the most prominent pitch in a \textit{gushe}, typically marked by increased duration; \textit{ist} (“stand”) is a cadential pitch that functions as an intermediate phrase-final note, distinct from the \textit{shāhed}; \textit{moteghayyer} (‘alterable’) relates to a pitch that may be replaced by another (either a quartertone or a semitone higher or lower) during the course of the gushe~ \cite {W-15}.}) associated with each \textit{gushe}. By organizing these diverse perspectives in a standardized, comparative format, the dataset enables users to cross-reference these theoretical descriptions with actual performance data. This structure supports both verification of annotations and the broader goal of examining how theoretical frameworks align—or diverge—from performance practice.

\begin{figure*}[!ht]
  \centering
  \includegraphics[width=\textwidth]{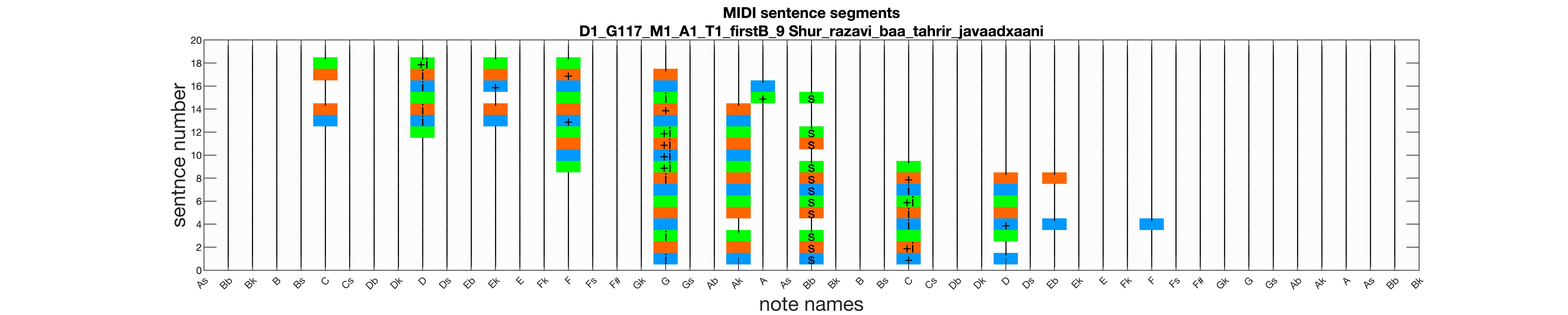}
  \caption{Note distribution and functional note annotations across symbolic segments in the piece \textit{Shur--Razavi bā tahrir-e javādxāni}. }
  \label{fig:segment-pitch-profile}
\end{figure*}

Figure~\ref{fig:segment-pitch-profile} visualizes the pitch content of symbolic segments extracted from the performance of the piece \textit{Shur--Razavi baa tahrir javaadxāni}. Each row corresponds to one sentence-level segment, and the vertical placement of colored blocks indicates the set of notes used in that segment. While the colors serve only to visually separate adjacent rows, annotated symbols convey musically meaningful roles: the central pitch or \textit{shāhed} is marked with \textbf{s}, the final note of each segment with \textbf{i}, and the most prominent note---in terms of duration---with \textbf{+}. This representation enables a visual analysis of modal focus and functional note behavior over the course of the performance. It also facilitates direct comparison with available theoretical descriptions of pitch roles and gushe structure in the radif tradition.

\begin{table*}[ht]
\centering
\caption{Completed and In-Progress Repertoires in the IRMA Dataset. \textit{Note: Audio recordings and third-party transcriptions referenced here are not included in the public release due to copyright.}}

\label{tab:radif_repertoires}
\begin{tabular}{p{3.5cm} c c p{7.5cm}}
\toprule
\textbf{Radif Source}  & \textbf{\# of Pieces} & \textbf{\# Dastgāhs/\={A}vāzes} & \textbf{Available Materials} \\
\midrule
Karimi (vocal) & 145 & 12 & Audio recordings; three-level symbolic MIDI files; PDF and Finale transcriptions; pitch and energy data tables; and comprehensive computational analyses (see Section~\ref{Karimi}). \\
\addlinespace
Karimi (vocal, 2nd version) & 38 & 12 & Audio recordings; pitch and energy data tables; automatically generated MIDI and MusicXML files for all pieces; sentence-level alignment of audio and MIDI (under review). \\
\addlinespace
Davāmi (vocal) & 188 & 13 & Audio recordings; pitch and energy data tables; three-level MIDI, PDF, and Finale transcriptions. \\
\addlinespace
Mirzā Abdollāh (instrumental) & 249 & 13 & Multiple audio recordings per piece; one MIDI file, one PDF transcription, and one Finale file per piece. \\
\addlinespace
Shajariān (vocal, unofficial recordings) & 131 & 12 & Unofficial audio recordings from Shajariān's teaching sessions; pitch and energy data tables; one symbolic MIDI file, one PDF transcription, and one MusicXML file per piece; sentence-level audio–MIDI alignment (under review). \\
\addlinespace
Kāzemi (vocal) & 339 & 13 & One audio recording per piece; pitch and energy data tables. MIDI and transcription materials are under development. \\
\addlinespace
\bottomrule
Tahrir Collection (various vocalists) & >260 & 12 & One audio recording per piece; pitch and energy data tables; MIDI and Finale transcriptions; sentence-level audio–MIDI alignment (under review). \\
\bottomrule
\end{tabular}
\end{table*}

The symbolic transcriptions, pitch and energy data, alignment files, and computational analyses included in the IRMA dataset are licensed under the \href{https://creativecommons.org/licenses/by-nc/4.0/}{Creative Commons Attribution–NonCommercial 4.0 International License (CC BY-NC 4.0)}. The dataset is publicly hosted at \textit {\url{https://github.com/SepiSha/irma-dataset}}. Due to copyright restrictions, the original audio recordings as well as transcriptions authored by other masters (e.g., Talāyi, Pāyvar) are referenced but not redistributed in this release. We are actively seeking appropriate permissions to include such materials in future versions of the dataset.

\section{Data Architecture and Encoding}

\subsection{Unique Identifier Code}
\label{code}
Throughout this work, we refer to individual pieces (\textit{gushes}) across various code components and documentation. For each \textit{gushe}, the IRMA dataset includes at least one audio file, one MIDI file, and one PDF transcription, drawn from editions of the radif by different musicians and scholars. In some cases, multiple audio files are available, reflecting different performances or transcription sources. For example, the vocal radif of Karimi is represented by three distinct transcriptions: Masoudieh (2004/1997) \cite{M1}, Atrāyi (2003), and Tahmāsbi (1995) with subtle variations. Similarly, the instrumental radif of Mirzā Abdollāh includes performances of each gushe by Alizāde (tār), Talāyi (tār and setār), Jamshid Andalibi (ney), and, in some cases, vocal renditions by Mohsen Kerāmati. To uniquely reference each musical unit, we introduce an identifier code which is explained in Section \ref{code}. 

Each identifier code should include the following information:

\begin{itemize}
  \item The position of the specific \textit{gushe} within various transcribed versions of the \textit{radif}.
  \item The name of the performer and the instrument used in the recording.
\end{itemize}

To generate consistent identifier codes, we first compiled a comparative table of gushes across all available published radifs. Although this study does not analyze every radif in detail, one of our primary goals in constructing the database was to create an inclusive and extensible structure that would accommodate future expansions. Our initial naming convention is based on the existing gushe names found in 14 different radifs. Once the musical content of these gushes is computationally analyzed, the naming system can be refined—for example, by merging variant names that refer to the same musical material or distinguishing identically named gushes with divergent content. 

\begin{figure}[h]

  \centering
  \includegraphics[width=\linewidth]{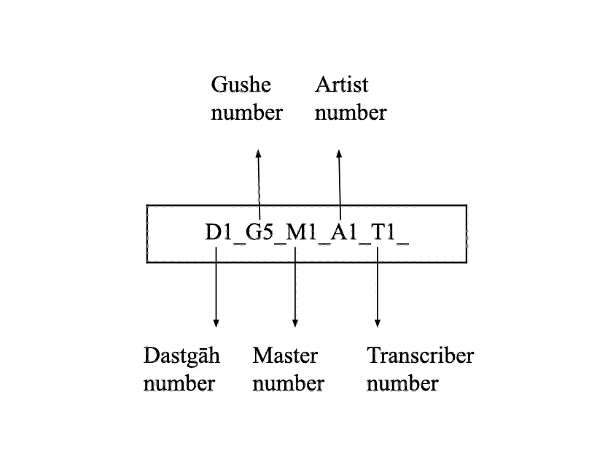}
  \caption{The Identifier Code in the Database
}
  \label{idcode}
  \Description{The Identifier Code in the Database
}
\end{figure}

Figure \ref{idcode} shows the format of the identifier codes in our database.

The parts of each identifier code in the IRMA Dataset are separated by underscores. To interpret the numeric values in each part of the code, a set of reference tables is required. For example, the \texttt{D}-part corresponds to the \textit{dastgāh} number, which is defined in a fixed table containing 13 rows—one for each \textit{dastgāh}. Similarly, the \texttt{G}-part refers to the \textit{gushe} number, which is determined by a comprehensive Excel table in the dataset containing 1,352 lines across 14 selected transcriptions.

While the \textit{dastgāh} table is fixed within the IRMA Dataset, the other tables (e.g., for performer codes, transcription types, etc.) are designed to be extensible, allowing users to add new rows to accommodate future developments or additional sources.

\subsection{Transcription Types and Audio-MIDI Alignment Using Dynamic Time Warping}

When we began this project in 2016, no existing MIDI dataset of any \textit{radif} was available that included pitch bends or other articulation-specific details necessary for computational analysis. Since Iranian music employs neutral intervals and nuanced intonational inflections, pitch bends are essential for accurately capturing melodic detail and expressive ornamentation. The dataset we introduce here is the first of its kind to address this gap. It incorporates detailed theoretical and musicological annotations for both the audio recordings and the symbolic transcriptions of each \textit{gushe} and phrase. 

The foundation of our analytical framework is based on Karimi’s performance of the vocal \textit{radif}, paired with Masoudieh’s transcription~ \cite{M1}. For Karimi’s \textit{radif}, we provide three types of transcriptions, as described below:

\emph{Type A:} The original transcription by Masoudieh of Karimi’s vocal radif.

\emph{Type B:} A digital rendering of Masoudieh’s transcription, typeset in Finale. This version includes interpretive additions such as markings for tekye (ornamental emphasis) and pitch bends, and is exported in both MIDI and PDF formats. Minor corrections are applied where necessary; these may vary depending on the transcriber. In the case of Masoudieh’s transcription, only a few typographical errors were corrected, and any changes are highlighted in red within the primary score.

\emph{Type C:}  A performance-informed transcription created by comparing the MIDI files from Type B with the corresponding audio recordings using our Radif Toolbox\footnote{Radif Toolbox is a collection of MATLAB scripts originally developed as part of the first author’s dissertation. It is designed to extract musicological information such as interval patterns, pitch drift, audio–MIDI alignment, sentence-by-sentence analysis of range and shāhed, and histogram representations of both MIDI and audio data. The toolbox was published as a GitHub repository in 2021 and has since been translated into Python for further development and expansion:  \url{https://github.com/SepiSha/radifToolBox}}. Adjustments are made so that the transcription reflects the nuances of the performance as closely as possible. This version is intended primarily for musicological analysis and emphasizes fidelity to the recorded interpretation.

\begin{figure}[h]

  \centering
  \includegraphics[width=\linewidth]{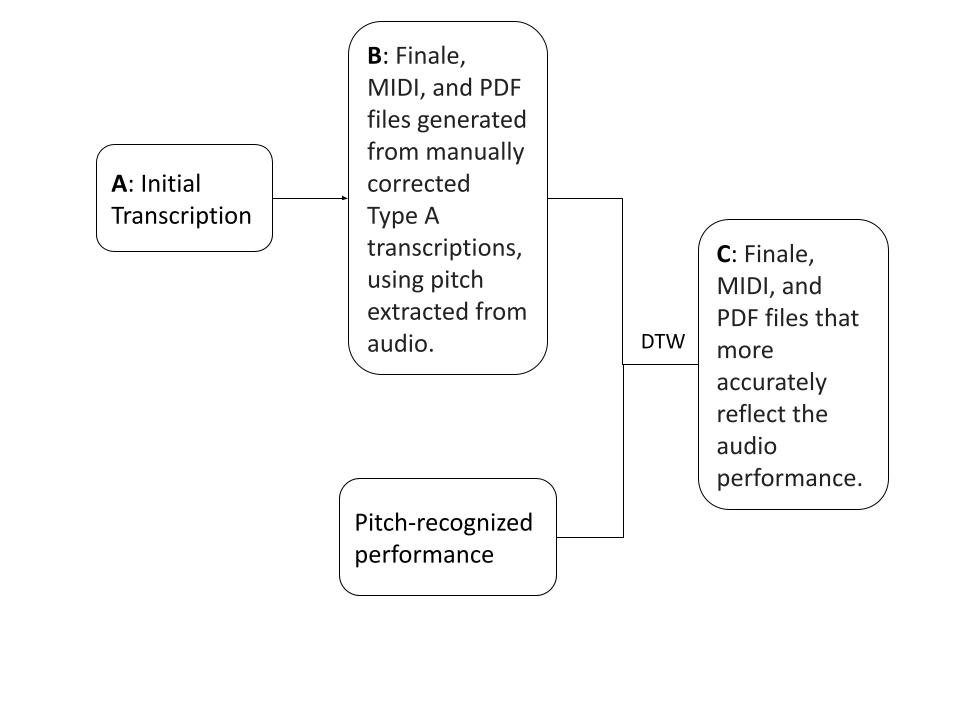}
  \caption{Comparison of Transcription Types: Notational vs. Performance-Informed Layers}
  \label{Types}
  \Description{Transcription types A, B, and C
}
\end{figure}

Figure~\ref{Types} illustrates the distinctions among these three types of transcriptions in the radif database.

\subsection{Encoded Features and Analysis}

Part of the development of this dataset involves selecting appropriate musical and analytical features for each \textit{gushe}, informed by both scholarly studies of the \textit{radif} and empirical observations of performance practice. Among the various characteristics, particular emphasis is placed on the automatic identification of note functions—such as \textit{shāhed} and \textit{ist} —as well as the detection of pitch intervals used in performance, the melodic range of each \textit{gushe}, and its internal phrase structure.%

Once the relevant features are selected, each gushe can be represented as an $n$-dimensional vector, where $n$ is the number of extracted features. These features may be weighted according to their relative importance, and the resulting vectors are organized into a matrix structure. This matrix forms the basis for further analysis using data mining and machine learning techniques.
 
One of the main challenges in applying computational tools to musical data analysis is the potential loss of interpretability. While advanced mathematical techniques can produce powerful results, they often yield outputs that are difficult to explain within a musicological context. To address this, we adopt a multi-layered approach that incorporates different levels of abstraction at each stage of the analysis. For instance, we bridge the gap between working solely with audio recordings and solely with transcriptions by aligning the two throughout the process. This allows us to construct a symbolic database that more closely reflects performance rather than relying entirely on notated sources. 

\section{Karimi’s Radif as Foundational Core}
\label{Karimi}

The development of the IRMA Dataset began with a detailed musicological and computational analysis of Karimi’s vocal \textit{radif}. Karimi is one of the most prominent vocalists of the 20\textsuperscript{th} century, and his complete repertoire was recorded in high quality and transcribed by the renowned musicologist Masoudieh~\cite{M1}. This initial corpus served as the foundation for the dataset's structure, design, and feature extraction framework. After establishing a robust structure based on Karimi’s repertoire, we incrementally incorporated additional materials, including the MIDI files of Davāmi’s vocal \textit{radif} and various instrumental \textit{radifs}.

In the dataset folder corresponding to Karimi’s \textit{radif}, each \textit{dastgāh} is organized as a separate subfolder. Within each \textit{dastgāh} folder, the following structure is used:

\begin{itemize}
    \item \texttt{performance}: Contains one MP3 file and two CSV files for each \textit{gushe}—one for pitch data and one for energy (RMS) information. Pitch tracking was performed using the pYIN algorithm via Sonic Annotator~\cite{MD-8}.

    \item \texttt{Masoudieh\_transcription}: Includes PDF files of Masoudieh’s original transcriptions for the \textit{gushes} in each \textit{dastgāh}.

    \item \texttt{typeB\_transcription}: Provides MUSX (Finale), MIDI, and PDF files for each \textit{gushe}. Each \textit{gushe} is also accompanied by a text file containing detailed timing and pitch information, as well as MIDI duration values based on audio alignment.

    \item \texttt{typeC\_transcription}: Similar to Type B, this folder includes MUSX, MIDI, and PDF files, along with note-level timing and pitch data derived directly from the audio using our alignment tools.

    \item \texttt{alignment\_images}: Contains subfolders for each \textit{gushe}, with sentence-level visual alignments between the audio and MIDI for both Type B and Type C transcriptions. It also includes segmented images from the original Type A transcription.

    \item \texttt{histograms}: Includes various visualizations for each \textit{gushe}, such as pitch histograms (audio and MIDI), pitch drift diagrams, and sentence-level representations of note events.
\end{itemize}


\section{Culture-Specific Considerations in Dataset Design}

\begin{table*}[!ht]
\centering
  \caption{ order of āvāzes of Shur in 13 selected transcriptions}
  \label{tab:orders}
  \begin{tabular}{ccccccl}
    \toprule
    1.Sabā- Pāyvar: dastur santur & 2. Borumand-Talāyi & 3. Borumand-During & 4. Karimi-Tahmāsbi & 5. Karimi-Masoudieh & 6. Karimi-Atrāyi \\
    \midrule
    Shur & Shur	& Shur & Shur & Shur & Shur\\
    Bayāt-tork & Bayāt-kord & Abuatā	 & Abuatā	 & Abuatā	 & Bayāt-kord\\
    Afshāri	& Bayāt-tork & Afshāri & Afshāri & Dashti & Abuatā\\
    Dashti	& Abuatā	& Dashti	& Dashti & Afshāri & Afshāri \\
    & Afshāri & Bayāt-kord &	&		 &Dashti\\
  \bottomrule
\end{tabular}
\end{table*}

\begin{table*}
\centering

  \label{tab:freq}
  \begin{tabular}{cccccccl}
    \toprule
    7.Davāmi& 8. Shahnāzi & 9. Marufi & 10. Habib Somāi & 11. Mirzā Abdollāh  & 12. Pāyvar (dore ebtedāyi) & 13. Montazem-ol-hokamā \\
    \midrule
    Shur & Shur	& Shur & Shur & Shur & Shur & Shur \\
    Abuatā	& Abuatā	& Abuatā	& Abuatā	& Bayāt-kord & Abuatā & Afshāri\\
    Afshāri	& Dashti	& Dashti	& Dashti	& Abuatā	& Bayāt-tork & Bayāt-kord\\
    Dashti	& Afshāri	& Afshāri	& Bayāt-tork& Dashti	 & Afshāri	& Bayāt-tork \\
    Bayāt-kord & Bayāt-tork & Bayāt-tork & & 		Afshāri&	Dashti&	Dashti\\
    Bayāt-tork& & & &				Bayāt-tork	& & \\
  \bottomrule
\end{tabular}

\end{table*}

The structure of the \textit{radif} tradition presents a number of challenges that are deeply rooted in oral transmission, performer-specific variation, and the lack of standardized notation. These characteristics pose significant obstacles for computational modeling, especially when aiming for consistency across symbolic encoding, alignment procedures, and corpus-level analysis. In the context of the IRMA Dataset, addressing these cultural and historical complexities is essential for designing representations that remain faithful to musical practice while enabling scalable, algorithmic processing. Below, we outline key issues encountered during dataset development that reflect these constraints. For example, several challenges arose in establishing a unified naming convention:

- The same gushe may appear under different names in various radifs. For instance, in the māhur section of Karimi’s vocal radif, two distinct gushes appear under the names tusi and nasirkhāni. In contrast, the instrumental radif of Borumand contains a single gushe labeled nasirkhāni yā tusi (“nasirkhāni or tusi”), suggesting that the two names may refer to the same musical content. In terms of melodic material, tusi in Karimi closely resembles nasirkhāni yā tusi in Mirzā Abdollāh’s radif, whereas nasirkhāni in Karimi more closely aligns with a different gushe in Mirzā Abdollāh’s radif, known as chahārpāre yā morādkhāni.

- Some \textit{gushes} appear under slightly different names across various \textit{radifs}, which poses significant challenges for automating the identification and comparison of corresponding pieces. For example, \textit{shur-e bālā} in Marufi’s \textit{radif} corresponds to \textit{shur-e pāyin-daste} in Borumand’s version, as found in both the Talāyi and During transcriptions. This apparent contradiction arises from two different uses of directional terminology: one based on pitch height, and the other on physical fingerboard position. In some traditions, the term \textit{bālā} (“upper”) refers to the higher register of a \textit{gushe} in terms of pitch. In contrast, in the context of setār and tār technique, \textit{pāyin-daste} (“lower hand position”) refers to the part of the fingerboard closer to the tuning pegs, which is used to play higher pitches. Conversely, \textit{bālā-daste} (“upper hand position”) refers to the section of the neck further from the nut, which produces lower pitches. This inversion between physical location and acoustic register can lead to inconsistencies in naming across different transcriptions, complicating symbolic alignment and automated matching in the dataset.

- Some gushes appear in only a subset of radifs, while being absent from others. For instance, gushe-ye Baghdādi is the thirteenth gushe in the Abuatā section of Musā Marufi’s radif, but it does not appear in the radifs of Mirzā Abdollāh, Karimi, or Davāmi.

- Some gushes share the same name across different radifs but differ significantly in melodic content and range. For example, the darāmad of māhur in Mirzā Abdollāh’s radif differs notably from the darāmad of māhur in Shahnāzi’s radif, both in terms of melodic structure and register.

- Different transcriptions of the same radif may contain varying numbers of gushes. For instance, Atrāyi’s transcription of Karimi’s radif includes more gushes than the version transcribed by Masoudieh. Atrāyi, a direct student of Karimi, transcribed the radif based on in-person lessons at the Tehran Conservatory of Music (Honarestān), whereas Masoudieh’s version is based on a recorded performance.

- In some cases, a radif includes multiple occurrences of a gushe with the same name within a single dastgāh. For example, in the rāst-panjgāh of Neydāvud, the gushe parvāne appears twice: once as the ninth gushe following rāst, and again as the sixteenth following mobarqa. In such instances, we must determine whether these refer to a single musical unit or distinct entities, and accordingly assign either a shared identifier code or two separate ones.

- Minor variations in gushe names across sources may be attributed to typographical errors or mispronunciations by performers. For instance, in the segāh and chahārgāh sections of Neydāvud’s radif, the gushe name khazān appears, which may be a misreading of hozzān due to the similarity of letters in the Persian script. Similarly, in the homāyun section of the same radif, the name farhang is listed, which may have been intended as farang.

- Some performers and transcribers separate certain \textit{foruds} (cadential phrases) and \textit{chahārmezrābs} (metrical instrumental pieces) from adjacent \textit{gushes}, assigning them distinct names and index numbers. Even within a single \textit{radif}, this practice is applied inconsistently, presenting challenges for standardization. Addressing these inconsistencies requires careful consideration in the design of the dataset structure and identifier schema.

- The ordering of dastgāhs, āvāzes, and gushes varies across different radifs, as illustrated in Table \ref{tab:orders}. However, certain clusters of gushes—or “islands of consistency”—retain a relatively stable sequence across most sources, providing reference points for comparison and alignment.

Table~\ref{tab:orders} presents the sequence of āvāzes within Shur across 13 selected transcriptions and notations of vocal and instrumental radifs. In the table, each entry in the first row lists the radif narrator (rāvi) followed by the transcriber, separated by a hyphen.\footnote{Davāmi is transcribed by Pāyvar; Habib Somāi by Talieh Kāmrān; and Mirzā Abdollāh by Kiyāni.}

\section{Musicological Applications}

The symbolic–audio alignment provided in the IRMA Dataset enables a range of pedagogical and computational applications. For educators and students of Iranian classical music, sentence-level visualizations of aligned transcriptions and performances offer an intuitive way to study phrasing, ornamentation, and intonation in context. For MIR tools, these alignments provide high-quality ground truth for tasks such as automatic transcription, onset detection, and expressive feature extraction. By tightly linking symbolic notation with performance audio, the dataset facilitates the development and evaluation of culturally informed MIR systems.

One of the key applications of this dataset is the analysis of microtonal intervals in Persian music, grounded directly in performance data. Because the symbolic transcriptions are aligned with audio recordings and enriched with pitch-bend information, the dataset enables precise extraction of pitch content as it is actually realized by performers. This supports the identification of microtonal intervals—such as neutral seconds, quartertones, and subtle inflections—that are central to the expressive vocabulary of Iranian classical music but often absent or underrepresented in notated sources. By analyzing the intervallic relationships between consecutive notes across different \textit{gushes}, the dataset supports empirical studies of intonation, tuning tendencies, and stylistic variation among vocalists and instrumentalists. These findings offer a performance-based foundation for modeling Iranian modal systems with greater cultural and acoustic specificity.

Another central application is the analysis of vocal ornamentation, particularly \textit{tahrir}, which plays a defining role in Iranian classical singing. The dedicated \textit{tahrir} section of the dataset includes over one hundred annotated examples drawn from historical recordings of prominent 20\textsuperscript{th}-century vocalists. Each sample is aligned with symbolic transcriptions and MIDI representations, enabling the study of melodic, rhythmic, and expressive characteristics of \textit{tahrir} gestures in both qualitative and quantitative terms. This resource supports detailed analysis of pitch trajectories, oscillation rates, temporal patterns, and contextual usage, facilitating cross-singer comparisons and stylistic classification. It also lays the groundwork for developing computational models of vocal ornamentation, contributing to both performance analysis and MIR applications such as ornament detection, synthesis, and transcription.

\section{Conclusion and Future Work}

The IRMA Dataset provides a computationally structured and musically grounded resource for the study of Iranian classical music, with particular attention to the \textit{radif} tradition. By integrating aligned audio and symbolic data, pitch-bend information, phrase-level annotations, and theoretical mappings, IRMA addresses major gaps in existing corpora—particularly in the representation of microtonal intervals, vocal ornamentation, and performance variation.

Future development will focus on expanding the dataset to include additional \textit{radif} lineages, performers, and instrumentation, as well as increasing coverage of underrepresented musicians and repertoires. Ongoing efforts include refining the identifier system and standardizing metadata schemas to improve interoperability and scalability. We also aim to support downstream applications such as automatic transcription, modal classification, expressive feature modeling, and cross-cultural comparative analysis.

Beyond its technical applications, IRMA contributes to the broader goals of cultural heritage preservation and musicological accessibility. It offers a platform that bridges traditional scholarship with contemporary computational methodologies. We welcome collaboration from researchers, musicians, and digital humanists to shape the continued evolution of the dataset.

\begin{acks}

The development of the IRMA dataset began as part of the first author's doctoral dissertation at the Graduate Center, City University of New York, under the supervision of Professor Stephen Blum. The authors gratefully acknowledge his guidance, encouragement, and deep insight throughout the early stages of this project. They also thank Joel Rodriguez Caraballo, who contributed to the project as music engraver and patiently endured nearly a decade of revisions and corrections.

\end{acks}

\bibliographystyle{ACM-Reference-Format}
\bibliography{references}

\end{document}